\documentclass[pdflatex,sn-nature]{sn-jnl}% Style for submissions to Nature Portfolio journals
\usepackage{graphicx}%
\usepackage{multirow}%
\usepackage{amsmath,amssymb,amsfonts}%
\usepackage{amsthm}%
\usepackage{mathrsfs}%
\usepackage[title]{appendix}%
\usepackage{xcolor}%
\usepackage{textcomp}%
\usepackage{manyfoot}%
\usepackage{booktabs}%
\usepackage{algorithm}%
\usepackage{algorithmicx}%
\usepackage{algpseudocode}%
\usepackage{listings}%
\theoremstyle{thmstyleone}%
\theoremstyle{thmstyletwo}%

\theoremstyle{thmstylethree}%

\begin{document}

\title{Atmospheric asymmetries in WASP-121\,b revealed by rotational transits detected with JWST}

\author*[1,2]{\fnm{Cyril} \sur{Gapp}}\email{gapp@mpia.de}

\author[3]{\fnm{Aurélien} \sur{Falco}}

\author[1,4]{\fnm{Thomas M.} \sur{Evans-Soma}}

\author[5,6]{\fnm{David K.} \sur{Sing}}

\author[7]{\fnm{Shashank} \sur{Dholakia}}

\author[8]{\fnm{Vivien} \sur{Parmentier}}

\author[9]{\fnm{Jérémy} \sur{Leconte}}

\author[1]{\fnm{Eva-Maria} \sur{Ahrer}}

\author[6]{\fnm{Guangwei} \sur{Fu}}

\affil[1]{\orgname{Max-Planck-Institut f\"{u}r Astronomie}, \orgaddress{\street{K\"{o}nigstuhl 17}, \city{Heidelberg}, \postcode{D-69117}, \country{Germany}}}

\affil[2]{\orgdiv{Department of Physics and Astronomy}, \orgname{Heidelberg University}, \orgaddress{\street{Im Neuenheimer Feld 226}, \city{Heidelberg}, \postcode{D-69120}, \country{Germany}}}

\affil[3]{\orgdiv{Université Côte d'Azur}, \orgname{Observatoire de la Côte d'Azur, CNRS}, \orgaddress{\street{Bd de l'Observatoire, CS 34229}, \city{Nice}, \postcode{06304}, \country{France}}}

\affil[4]{\orgdiv{School of Science}, \orgname{University of Newcastle}, \orgaddress{\city{Callaghan}, \state{NSW}, \country{Australia}}}

\affil[5]{\orgdiv{Department of Earth \& Planetary Sciences}, \orgname{Johns Hopkins University}, \orgaddress{\city{Baltimore}, \state{MD}, \country{USA}}}

\affil[6]{\orgdiv{Department of Physics \& Astronomy}, \orgname{Johns Hopkins University}, \orgaddress{\city{Baltimore}, \state{MD}, \country{USA}}}

\affil[7]{\orgdiv{School of Mathematics and Physics}, \orgname{The University of Queensland}, \orgaddress{\city{St Lucia}, \state{QLD 4072}, \country{Australia}}}

\affil[8]{\orgdiv{Laboratoire Lagrange, Observatoire de la Côte d’Azur}, \orgname{Université Côte d’Azur}, \orgaddress{\city{Nice}, \postcode{06903}, \country{France}}}

\affil[9]{\orgdiv{Laboratoire d’Astrophysique de Bordeaux}, \orgname{Univ. Bordeaux}, \orgaddress{\street{Allée Geoffroy Saint-Hilaire}, \city{Pessac}, \postcode{33615}, \country{France}}}

\abstract{Close-in exoplanets are tidally locked to their host star and thus exhibit extreme atmospheric temperature gradients. It has been theorized that the fraction of star light absorbed by such planets during transit changes as a function of orbital phase as progressively hotter or colder atmospheric gas rotates into view, but this effect has not been observed so far. Here, we show that two transits of the ultra-hot Jupiter WASP-121\,b, acquired with JWST/NIRSpec and NIRISS, exhibit asymmetric transit light curves caused by the planet's rotation during transit. We observe increasing CO absorption and slightly decreasing H$_2$O absorption in the transmission spectrum, as the planet rotates. These results are indicative of a stronger longitudinal temperature gradient across the evening than across the morning terminator, consistent with higher temperatures in the eastern half than in the western half of the dayside. The observed changes of the transmission spectrum with orbital phase are in line with the temperature increase causing thermal dissociation of H$_2$O, while CO remains abundant. The observation of longitudinal gradients of atmospheric temperature and chemistry from the planet's rotational transit provides a new probe for constraining atmospheric heterogeneity using JWST beyond differences between morning and evening terminators from limb asymmetries.}


\maketitle

\section*{Main text}
One of the most extensively used observational techniques to study the atmospheres of exoplanets is transmission spectroscopy, the study of how absorption in an exoplanet's atmosphere modifies star light transmitting through it while the planet transits in front of its host star \cite{seager00}. The resulting transmission spectrum, the planet's absorbing cross-sectional area relative to its star's disk area as a function of wavelength, probes the planet's atmosphere in an annulus around its limb. The longitudinal extend of this annulus around the equator (the `opening angle') varies from single degrees in cold planets to tens of degrees in hot ones \cite{caldas19}. Since exoplanets rotate while they transit, the region in their atmospheres that is being probed at any point during their transits steadily changes in longitude as the planet's orbital phase changes. Subsequently, their transmission spectra can change as a function of phase during transit. These changes are observable through phase-dependent changes of the planet's absorbing cross-sectional area \cite{falco24}, which we refer to as rotational transits. Thus, measuring rotational transits would open up the possibility to constrain transiting exoplanets' atmospheric temperatures and chemical compositions as functions of longitude \cite{burrows10}.

Some of the hottest known exoplanets, the ultrahot Jupiters, gas giants with equilibrium temperatures exceeding $\sim 2000$\,K, rotate by more degrees during their transits than their opening angles \cite{wardenier22}. Thus, the star light transmitting through the planet at the end of the transit probes a longitudinal segment of the atmosphere that does not overlap with the segment probed at the transit's beginning. Therefore, ultrahot Jupiters are expected to exhibit strongly phase-dependent absorption signals in transit, delivering leverage into studying their atmospheric temperatures and chemical composition as functions of longitude. To that end, the longitudinal structures of these planets have been studied extensively with ground-based telescopes at high spectral resolution. In the ultrahot Jupiter WASP-76\,b, phase-dependent Fe absorption \cite{ehrenreich20,kesseli21,gandhi23} has revealed the planet's atmosphere to be asymmetric about the sub-stellar point. This asymmetry might be driven by Fe condensation in the nightside \cite{kesseli22,gandhi22}, cloud condensation around the leading limb \cite{savel22,gandhi22,maguire24}, higher temperatures in the trailing compared to the leading half of the planet \cite{wardenier21} or a combination of any of these factors. In WASP-121\,b, resolving the absorption signals of Na I, Fe I and H$\alpha$ which probe the planet's atmosphere at different heights has enabled to vertically and longitudinally resolve its atmospheric flow pattern with a deep day-to-night circulation and an equatorial jet in the upper atmosphere \cite{seidel25}.

These past achievements have been accomplished with spectroscopic observations of atoms and ions at optical wavelengths, where the absorption fraction of star light in the Earth's atmosphere is low. Similar observations at longer, infrared wavelengths targeting molecules suffer from increased telluric absorption, leading to distortions of the signals that can cause observations to not repeat in different observing nights \cite{wardenier24}. Thus, space-based observations in the infrared can compliment ground-based observations by revealing the longitudinal structure of molecular abundances in exoplanets from transit observations. From the ground, constraints on an exoplanet's three-dimensional atmospheric structure are derived from measuring Doppler shifts of absorption lines driven by the planet's rotation and winds in its atmosphere. These measurements are possible thanks to high spectral resolving powers that are not achievable in space. Therefore, measuring heterogeneity in exoplanet atmospheres using their rotations during transits observed with space telescopes requires analysis approaches that are distinct from line shift measurements from the ground.

\subsection*{Rotational light curve model}
At low spectral resolution, the Doppler shifts of spectral lines are lost, but the total fraction of star light absorbed by the planet can be measured very precisely by comparing the flux of star light during transit to the flux observed before and after the transit. When the planet rotates during transit, the fraction of absorbed star light can change as a function of time as different atmospheric temperatures and chemical compositions are probed \cite{falco24}. The analytical form of this function is unknown and thus, we choose a polynomial approximation which represents a Taylor expansion.

In the case of the ultrahot Jupiter WASP-121\,b, the three-dimensional atmospheric model SPARC/MITgcm \cite{showman09,parmentier18} predicts the planet to change its apparent radius due to its rotation during transit in a way that can be approximated well by a second-order polynomial (see Methods). Thus, we restrict the Taylor expansion of the planet-to-star radius ratio as a function of time to the second order. To measure these radius changes, we developed a rotational light curve model ($\mathcal{M}_{\rm{rot}}$) in which the exoplanet’s radius ($R_p$) relative to its host star's radius ($R_*$) is given by
\begin{equation}
    R_p/R_*(t) = \sum_{i=0}^2(t-t_0)^iR_i/R_*, \label{eq:Rp}
\end{equation}
where $t$ denotes time, $t_0$ is the time of inferior conjunction and $R_i/R_*$ are the polynomial coefficients. In this notation, $R_0/R_*$ represents the planet-to-star radius ratio at $t_0$, $R_1/R_*$ gives the planet's linear rate of radius change relative to the star's radius and $R_2/R_*$ gives the curvature of the radius variation in time. This light curve model is qualitatively different from past modeling approaches that adopted either a circle \cite{batman}, two stacked semi-circles \cite{catwoman} or arbitrarily complex cross-sectional shapes parameterized using Fourier series \cite{harmonica}. Since these models simulate transit light curves by displacing a time-independent planetary cross-sectional shape across the stellar disk, they neglect the changing viewing geometry of the planet during transit that changes its absorbing cross-sectional area as a function of time \cite{falco24}.

\subsection*{Results}
\begin{figure}[htbp!]
    \centering
    \includegraphics[width=\textwidth]{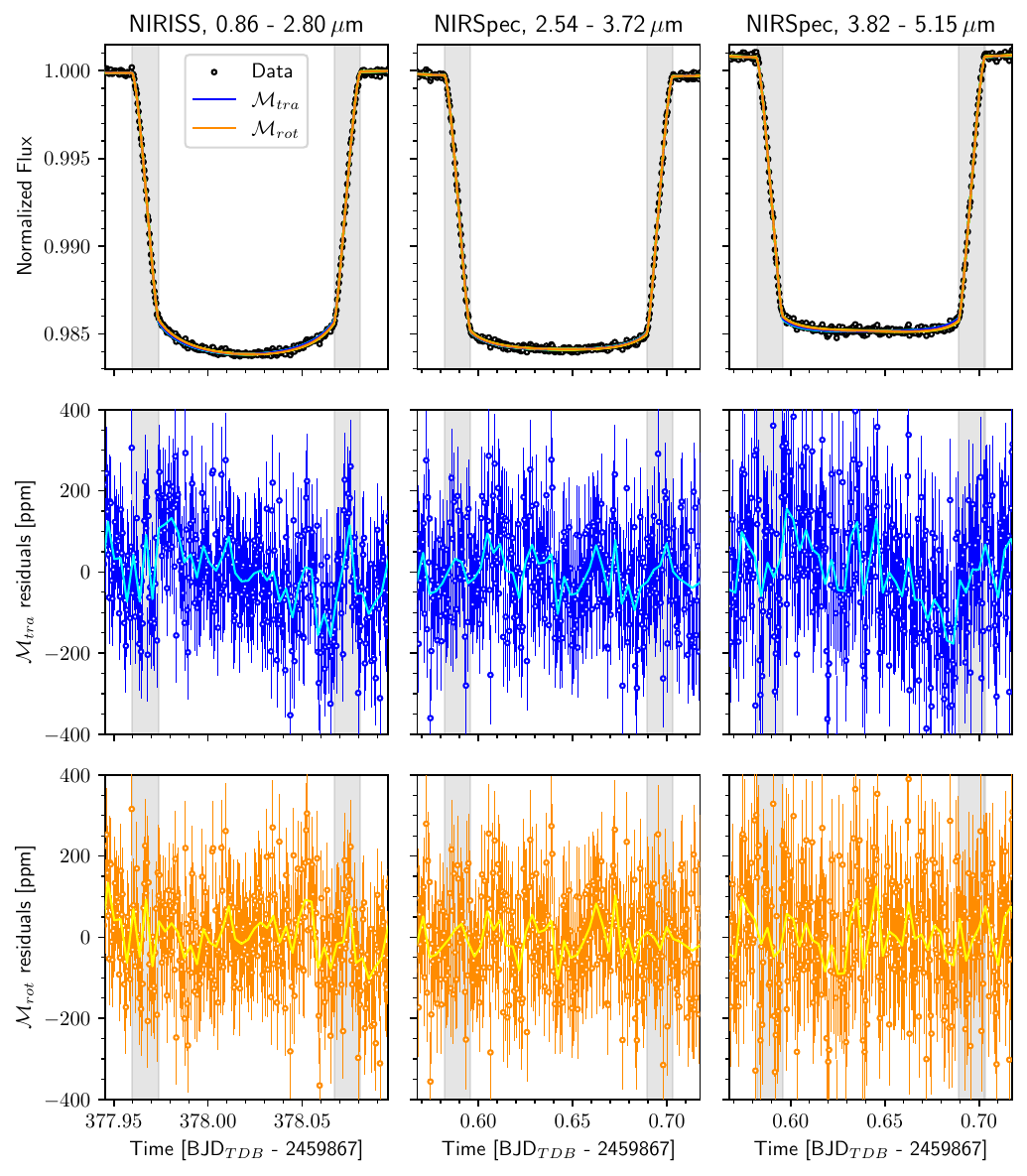}
    \caption{WASP-121\,b's white transit light curves observed with JWST/NIRISS SOSS and JWST/NIRSpec G395H. Both observations have been reduced with Firefly. The first row shows the raw light curves integrated over NIRISS's first grating order and NIRSpec's NRS1 and NRS2 detectors together with the maximum-likelihood translational ($\mathcal{M}_{\rm{tra}}$) and rotational ($\mathcal{M}_{\rm{rot}}$) models inferred from a Markov-Chain Monte Carlo (MCMC) sampling of the posterior probability. Circles depict the fluxes per integration normalized by the median fluxes measured during the secondary eclipse observed before the transit and error bars depict the $1\sigma$ intervals measured from 150,000 posteriors samples of the light curve (see Methods). Transparent cyan and yellow lines depict models calculated from 100 samples randomly drawn from the MCMC chains of $\mathcal{M}_{\rm{tra}}$ and $\mathcal{M}_{\rm{rot}}$, respectively. The second and third rows show the residuals between the maximum-likelihood models and the data. Vertical gray lines mark contact points 1, 2, 3, and 4 calculated from the planet's orbital parameters and radius \cite{winn10} with gray shaded regions indicating the times of ingress and egress.}
    \label{fig:white-lightcurve}
\end{figure}

We applied the rotational transit model to two transit observations of WASP-121\,b observed with the Near Infrared Spectrograph (NIRSpec \cite{MikalEvans23,EvansSoma25,Gapp25}) and the Near Infrared Imager and Slitless Spectrograph (NIRISS, \cite{splinter25,allart25}), respectively, onboard the James Webb Space Telescope (JWST). To quantify the statistical significances of the planet’s radius change during transit in these observations, we performed statistical hypothesis tests on the wavelength-integrated (‘white’) light curves with a translational light curve model ($\mathcal{M}_{\rm{tra}}$) that prescribes a time-invariant planet radius (i.e., $R_1/R_*=R_2/R_*=0$) as the null hypothesis and $\mathcal{M}_{\rm{rot}}$ with freely-fit $R_1/R_*$ and $R_2/R_*$ as the alternative. In a Bayesian framework, we estimate the ratio of the probabilities of the alternative and the null hypothesis using the Bayes factor ($B$) calculated by taking the ratio of the Bayesian evidence of $\mathcal{M}_{\rm{rot}}$ to the Bayesian evidence of $\mathcal{M}_{\rm{tra}}$. We ran this test on two independent data reductions of each observation to verify the results against reduction-level assumptions (see Methods). $\mathcal{M}_{\rm{rot}}$ is significantly preferred \cite{thorngren26} over $\mathcal{M}_{\rm{tra}}$ with $\ln(B)= 18.0$ and $\ln(B)=24.8$, respectively, in the two NIRISS SOSS data reductions (see Extended Data Table \ref{tab:evidence_niriss}) and $\ln(B)= 11.7$ and $\ln(B)=23.7$ in the two NIRSpec G395H reductions (see Extended Data Table \ref{tab:evidence_nirspec}). An inspection of the observations and models (see Fig. \ref{fig:white-lightcurve} and Extended Data Fig. \ref{fig:white-lightcurve_eureka}) suggests that the data between contact points 2 and 3 of the transit drive the detection of the planet’s rotation as the fit to the data using $\mathcal{M}_{\rm{tra}}$ results in systematic linear residuals in those data that are absent when $\mathcal{M}_{\rm{rot}}$ is applied. This demonstrates that the planet's rotation leads to it appearing larger in the second than in the first half of the transit which results in asymmetric light curve shapes in both observations. These asymmetric light curves cannot be adequately fit with a time-independent transit model that models the planet's cross-section using two semi-circles with a shared center point but different radii (so-called `asymmetric limbs') \cite{Gapp25}. Thus, the signal is dominated by the star light transmitting through different longitudes in the planet as a function of time as it rotates during the transit.

There is no evidence to fit for a free $R_2/R_*$ over setting $R_2/R_*=0$ in $\mathcal{M}_{\rm{rot}}$ likely due to a degeneracy of $R_2/R_*$ with the host star's limb darkening (see Extended Data Figs. \ref{fig:corner} and \ref{fig:radius-functions}). To examine the impact of setting $R_2/R_*=0$ in $\mathcal{M}_{\rm{rot}}$, we repeated the hypothesis test and found that the detections of WASP-121\,b's rotation-induced radius change remain significant in both observations with $\ln(B)=16.5$ and $\ln(B)=23.2$ in the NIRISS SOSS reductions (see Extended Data Table \ref{tab:evidence_niriss}) and $\ln(B)=13.6$ and $\ln(B)=24.9$ in the NIRSpec G395H reductions (see Extended Data Table \ref{tab:evidence_nirspec}). Another degeneracy between the rotation-induced radius change and the host star is driven by WASP-121\,A's gravity darkening caused by its rapid rotation: Data from the Transiting Exoplanet Survey Satellite (TESS) reveal that WASP-121\,b transits a brighter part of its host star's disk at the end of the transit than in the beginning, also creating an asymmetric light curve (see Extended Data Fig. \ref{fig:gravitydarkening}). WASP-121\,A's gravity darkening affects the NIRISS SOSS data, but is only a minor effect in the NIRSpec G395H data thanks to the latter instrument probing the transit in longer wavelengths (see Methods).

\begin{figure}[htbp!]
    \centering
    \includegraphics[width=\textwidth]{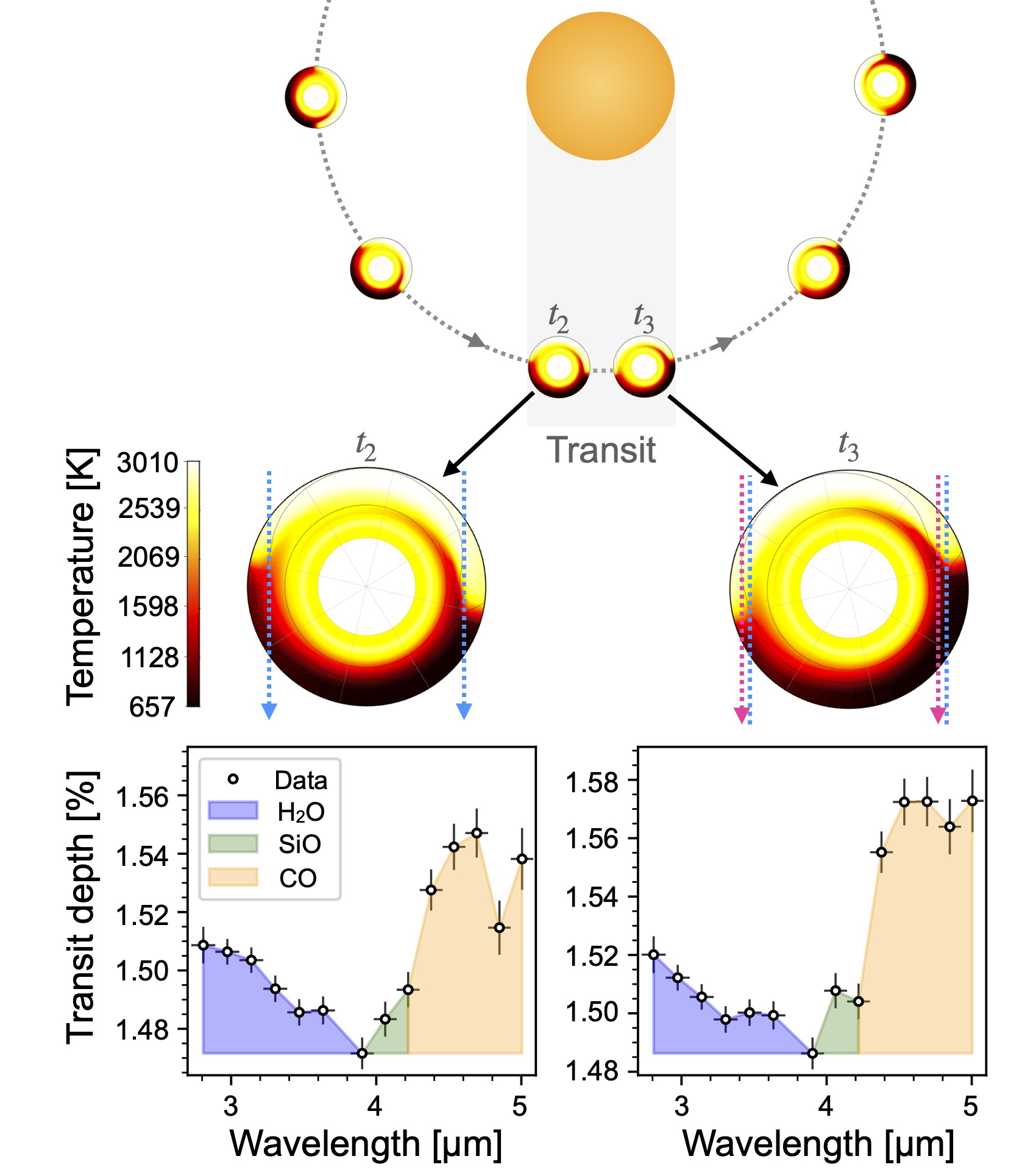}
    \caption{Visualization of WASP-121\,b's orbital motion and phase-dependent transmission spectrum. The transmission spectra at contact points two ($t_2$) and three ($t_3$) are shown in the lower row. The spectra were measured with NIRSpec G395H using the Firefly reduction and modeled adopting $R_2/R_*=0$. Circles depict medians and error bars indicate the $1\,\sigma$ range of 125,000 posterior samples. The planet's orbit and star's radius are shown to scale, but the planet's and its atmosphere's size are exaggerated for visual clarity. The color map on the planet depicts an equatorial cross-section of the planet's atmospheric temperature from the SPARC/MITgcm. The blue and magenta arrows in the zooms into the planet's temperature map indicate the approximate heights at which the atmosphere becomes opaque. In the $t_3$ map, dotted blue lines correspond to the heights probed at $t_2$ to demonstrate how the trailing limb grows while the leading limb shrinks as the planet rotates. In the transmission spectra, colored regions underneath the data points indicate which molecule is the dominant absorber in each wavelength channel \cite{Gapp25}.}
    \label{fig:visualization}
\end{figure}

Because of the degeneracy of $R_2/R_*$ with the host star's limb darkening and the contamination of the NIRISS SOSS data with the star's gravity darkening, we derive WASP-121\,b's transmission spectrum as a function of orbital phase from fits to the spectroscopic light curves of NIRSpec G395H with $\mathcal{M}_{\rm{rot}}$ adopting $R_2/R_*=0$. The polynomial coefficients as functions of wavelength in those data are shown in Extended Data Fig. \ref{fig:spectrum}. The transmission spectra at the orbital phases at the beginning and end of the transit (see Fig. \ref{fig:visualization}) firstly reveal that the fraction of star light absorbed by the planet increases as a function of time through the entire bandpass. Secondly, the CO absorption feature between $4.3\,\mu$m and $5.2\,\mu$m grows by $\sim 200$\,ppm relative to the rest of the spectrum between contact points 2 and 3. This increase of CO absorption is observed with a consistent amplitude in the NIRISS SOSS data (see Extended Data Fig. \ref{fig:co-lightcurve}). And thirdly, the H$_2$O feature shortward of $3.8\,\mu$m remains approximately constant, possibly shrinking slightly. Additionally, the absorption from SiO that occurs between $4.0\,\mu$m to $4.3\,\mu$m appears to slightly increase, though not as strongly as the CO absorption. With the narrow wavelength range of the SiO absorption feature and thus few data, however, it is challenging to draw definite conclusions.

\subsection*{Discussion}
\subsubsection*{Thermal structure of WASP-121\,b}
WASP-121\,b's tide-locked rotation leads to the trailing limb probing longitudes closer to the substellar point and the leading limb probing longitudes further away from the substellar point as the transit progresses (see Fig. \ref{fig:visualization}). Since both limbs' vertical extents are primarily set by the local temperature due to thermal expansion and the thermal dissociation of molecules, the leading limb becomes smaller and the trailing limb becomes larger throughout the transit. Therefore, measuring the temporal change of the fraction of absorbed star light throughout the transit yields insight into the longitudinal temperature gradients across the morning and the evening terminators.

In WASP-121\,b, the fraction of absorbed star light increases throughout the whole spectrum observed with NIRSpec G395H (see Fig. \ref{fig:visualization}.) This means that the longitudinal temperature gradient across the evening terminator that leads to an increase of the planet's cross-sectional area in time exceeds the temperature gradient across the morning terminator that causes a decrease of the total absorbing area. This observation is consistent with previous studies that suggested the eastern half of the dayside to be hotter than the western half \cite{MikalEvans23,seidel25} and results from the SPARC/MITgcm that also predicts WASP-121\,b's atmosphere to be asymmetric about the substellar point (see Extended Data Fig. \ref{fig:gcm-maps}).

\subsubsection*{Chemical gradients in WASP-121\,b}
We propose that the changes of WASP-121\,b's transmission spectrum throughout the transit are driven by the different distributions of molecules in WASP-121\,b's atmosphere: H$_2$O thermally dissociates on the dayside \cite{MikalEvans20,wardenier24,smith24,pelletier25}, leading to a higher abundance on the nightside, while CO is abundant in the entire atmosphere thanks to its stability against thermal dissociation \cite{parmentier18}. Thus, the progressively hotter gas rotating into view on the trailing limb becomes more and more depleted in H$_2$O, while the CO abundance remains approximately constant. Therefore, the increase of the transit depth is strong in the CO band owing to the increase in atmospheric scale height, but this effect is overwhelmed in the H$_2$O band due to thermal dissociation of H$_2$O as temperatures increase. The slight increase of SiO absorption as the transit progresses could indicate that SiO thermally dissociates less than H$_2$O, but the limited constraining power due to its narrow absorption feature makes the interpretation difficult.

\subsubsection*{Comparison to ground-based observations}
Phase-resolved observations of the H$\alpha$ line in WASP-121\,b obtained using the four-unit telescope (4-UT) mode of the Echelle Spectrograph for Rocky Exoplanets and Stable Spectroscopic Observations (ESPRESSO) at the Very Large Telescope (VLT) were used in a past study to trace the longitudinal variation of temperature in the uppermost part of the atmosphere where the planet's atmosphere escapes into space \cite{seidel25}. These data are sensitive to very low pressures in the atmosphere on the level of microbars where the atmosphere is extremely hot ($T>8000$\,K). The JWST observations presented here are sensitive to higher pressures than that around milibars, complimenting the existing temperature gradient constraints from high-resolution observations. Regarding chemical gradients in WASP-121\,b, phase-resolved atmospheric retrievals \cite{gandhi23} were carried out on ESPRESSO data obtained in its one-unit telescope (1-UT) mode \cite{borsa21}. The data are sensitive to atomic species such as Fe and did not reveal chemical gradients around the planet's terminators, but abundances consistent with homogeneous distributions \cite{gandhi23}. Since chemical gradients of atoms in exoplanets are driven by rain-out or ionization, the lack of atomic gradients is not in tension with the gradient in H$_2$O driven by thermal dissociation we observe with JWST.

Regarding longitudinal gradients of molecular abundances, the Doppler shifts of CO and H$_2$O lines in WASP-121\,b were previously measured as functions of orbital phase during three transits observed using the Immersion Grating Infrared Spectrometer (IGRINS) on the Gemini-South Telescope \cite{wardenier24}. The data revealed CO lines to become increasingly blueshifted during the transits indicating that CO absorption predominantly originates in the trailing limb that rotates toward the observer. The blueshift of H$_2$O lines, in contrast, decreased as the transits progressed, revealing H$_2$O absorption to be dominated by the leading limb that rotates away from the observer \cite{wardenier23}. These results are in line with the JWST observations presented here that also suggest CO, but not H$_2$O, absorption to trace the temperature increase in the trailing limb as the planet rotates during transit due to the thermal dissociation of H$_2$O. Thus, the consistency between the Gemini-South and JWST observations highlights the synergy between phase-resolved ground-based and space-based observations.

\subsubsection*{Future work}
\begin{figure}[htbp!]
    \centering
    \includegraphics[width=\textwidth]{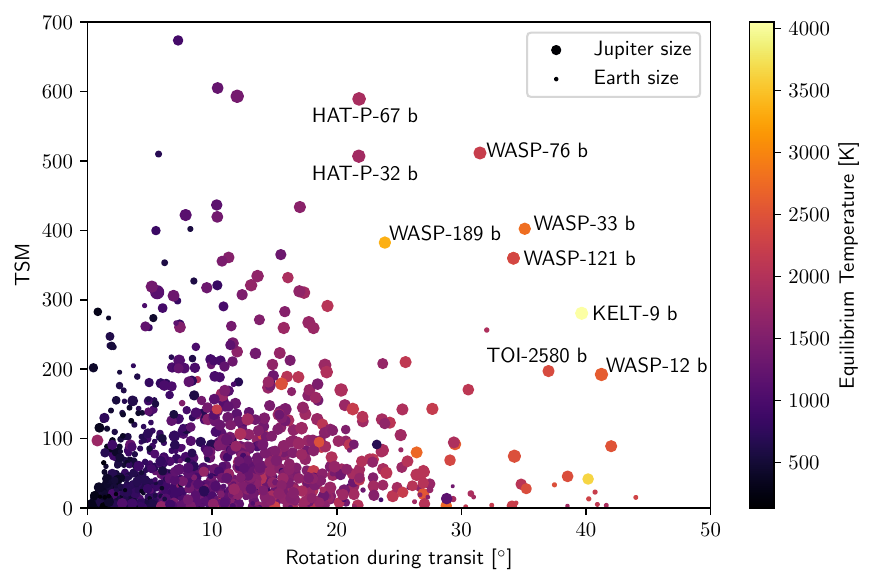}
    \caption{Known transiting exoplanets’ TSMs and rotations during transit estimated assuming synchronous rotation and zero eccentricity. Dot colors represent equilibrium temperatures and dot sizes are proportional to planet radii. The names of some of the planets that rotate the most during transit and are the most promising for atmospheric characterization using transmission spectroscopy are indicated. The data were taken from the NASA Exoplanet Archive \cite{exoplanetarchive}.}
    \label{fig:population}
\end{figure}

With the detection of WASP-121\,b's rotational transit and subsequently its phase-resolved transmission spectrum, the longitudinal gradients of temperature and chemistry around the planet's terminators become observable in transit at low spectral resolution. Beyond WASP-121\,b, the phase-dependent transmission spectra of several other exoplanets might be measurable with JWST. We propose to use exoplanets' rotations during transit, their equilibrium temperatures (delivering a proxy for their longitudinal temperature gradients), and their transmission spectroscopy metrics (TSM \cite{tsm}, giving an estimate of the signal-to-noise ratio of a planet's atmospheric features in transit) to guide the expectation for the signal amplitudes caused by rotational transits. Under these considerations, WASP-121\,b is one of the stand-out targets (see Fig. \ref{fig:population}). Some of the other promising targets, such as WASP-33\,b, WASP-189\,b and KELT-9\,b, orbit rapidly-rotating A-type stars for which gravity darkening has been shown to impact the planets' transit light curves by introducing asymmetric transits \cite{dholakia22,deline22,ahlers20}. Thus, constraints on these planets' rotational transits will require careful modeling of the stars' gravity darkening by combining transit light curves from different wavelength ranges as it was done in the present study using data from JWST and TESS.

\section*{Methods}
\subsection*{Observations and data reductions}
\subsubsection*{JWST/NIRSpec G395H observations}
The James Webb Space Telescope (JWST) observed the ultrahot Jupiter WASP-121\,b starting 145 minutes before the beginning of the planet's eclipse behind the star on October 14, 2022 (UTC), and finishing the observation 105 minutes after the end of the eclipse on October 15, 2022 (UTC). This observation was carried out as the telescope's observing program GO 1729 (PIs: Evans-Soma, Kataria) using the Near Infrared Spectrograph (NIRSpec) with the G395H grating and naturally included the transit of the planet in front of the star between these two eclipses. The observations were conducted using NIRSpec’s SUB2048 subarray and NRSRAPID readout pattern with 42 groups per integration and were reduced using the Fast Infrared Exoplanet Fitting Lightcurve (Firefly \cite{rustamkulov22,rustamkulov23,Sing2024Natur.630..831S}) and Eureka! \cite{eureka} pipelines. The details of the two data reductions performed with the two different pipelines are listed in ref. \cite{MikalEvans23,EvansSoma25}.

\subsubsection*{JWST/NIRISS SOSS observations}
JWST also observed a phase curve of WASP-121\,b using the Near Infrared Imager and Slitless Spectrograph (NIRISS) instrument starting before a planetary eclipse on October 10, 2023 (UTC) and ending 43.85 hours later after a second eclipse event. This observation was part of observing program GTO 1201 (PI: Lafrenière) and used the single object slitless spectroscopy (SOSS) mode with the SUBSTRIP256 subarray with 6 groups per integration.

The observations were reduced starting from the uncalibrated (\texttt{uncal.fits}) images using the Firefly reduction suite \citep{rustamkulov22,rustamkulov23,Sing2024Natur.630..831S}, which has been adapted for SOSS data \cite{Liu2025arXiv250408903L} with further details available in ref. \cite{Liu2025arXiv250408903L,Mukherjee2025arXiv250510910M,Wang26}. A notable recent addition is the removal of 1/f noise at the group-level stage, as described in ref. \cite{Radica2023MNRAS.524..835R}. From the cosmic ray, bad pixel, 1/f-removed, background-removed 2D images, we summed the white light curve using an aperture width of 34 pixels which produced a first order white light curve which minimized the light curve scatter, found to have a standard deviation of 110\,ppm and a median absolute deviation of 83\,ppm as characterized by a near-flat 100 integration segment shortly before transit.

We also performed a NIRISS SOSS reduction independent of the Firefly reduction using steps as outlined in ref. \cite{fu_water_22} (hereafter labeled the `Fu' reduction). It starts with the uncalibrated (\texttt{uncal.fits}) files, and we used the default JWST pipeline to produce the \texttt{rampfitsstep.fits}. Then, we did background and 1/f subtraction using pixels in the bottom 50 percentile in flux for column median subtraction and extracted the time series two-dimensional spectra. Based on the 100 points right before transit, we measured a standard deviation of 116\,ppm and a median absolute deviation of 88\,ppm.

\subsection*{JWST data analysis}
We approximate the rotation-induced change of the planet’s transit radius with a second-order polynomial in time (see main text). We implemented this light curve model using the Batman package \cite{batman}, in which the planet-to-star radius ratio is a constant. To calculate light curve models with time-dependent planet radii, we set up a Batman model for each data point of the time-series observation and set the planet-to-star radius ratio in the series of Batman models according to equation \ref{eq:Rp}.

To measure WASP-121\,b's phase-resolved transmission spectrum, we analyze a cut-out of the light curves that includes the data taken during both full eclipses and from 3.5 hours before to 3.5 hours after the transit mid-time output from both reductions of both the NIRSpec and NIRISS observations. Including the adjacent eclipses into the transit analysis is needed, because the planet's rotation around the transit results in a variation of its emitted flux as a function of time that would contaminate the transmission spectrum if it were not separated from the systematics' baseline \cite{Gapp25}. The full model we use to fit the observations is
\begin{equation}
    \mathcal{M}_{\rm{rot}} = (c_0 + c_1 t) \times \left(T(\theta)+P(p_0,p_1,p_2)\right), \label{eq:m_rot}
\end{equation}
where $(c_0 + c_1 t)$ is the stellar and instrumental baseline, $T(\theta)$ is the modified Batman model with transit parameters $\theta$, and
\begin{equation}
    P(p_0,p_1,p_2) =
    \begin{cases}
        0 & \text{during secondary eclipse} \\
        p_0 + p_1t + p_2 t^2 & \text{else} \label{eq:phasecurve}
    \end{cases}
\end{equation}
is an approximation of WASP-121\,b's partial phase curve around the transit. The translational light curve model $\mathcal{M}_{\rm{tra}}$ we use as a null hypothesis is identical to $\mathcal{M}_{\rm{rot}}$ apart from prescribing $R_1/R_*=R_2/R_*=0$.

\subsubsection*{White light curve fits}
In JWST/NIRSpec’s G395H observing mode, the incoming radiation gets dispersed over two detectors, NRS1 and NRS2. We integrated the time-series of stellar spectra from pixel column 300 to 2042 on the NRS1 and from pixel column 5 to 2010 on the NRS2 detectors to yield wavelength-integrated light curves for both detectors. The two light curves were then fit simultaneously. For the orbital parameters shared between both light curves, we adopted an eccentricity of zero, an argument of periapsis of $90^\circ$ and a period of 1.27492504 days \cite{bourrier2020a} and fit for the impact parameter ($b$) and semi-major axis normalized by the star’s radius ($a/R_*$). All other model parameters, the baseline coefficients $c_0$ and $c_1$ (Eq. \ref{eq:m_rot}), the partial phase curve parameters $p_0$, $p_1$ and $p_2$, the transit radius coefficients ($R_i/R_*$ in Eq. \ref{eq:Rp}), $t_0$ and the limb darkening coefficients were kept different between both light curves. For the limb darkening, we adopted a quadratic law with two free parameters ($u_1$ and $u_2$). The JWST/NIRISS white light curve was calculated by integrating the flux from detector column 33 to 2033 in the first grating order and fit in the same manner as the NIRSpec white light curve, but without the need to fit two independent light curves jointly.

We first fit the data from both instruments with a least-squares algorithm implemented into the lmfit package \cite{lmfit}. Then, we explored all fit parameters’ posterior probability distributions using the Markov chain Monte Carlo (MCMC) package emcee \cite{emcee} and estimated the models’ Bayesian evidences using the nested sampling package dynesty \cite{dynesty}. In both sampling approaches, we included systematic noise terms ($\sigma_\text{sys}$) we added to the uncertainties of each integration ($\sigma_{p,i}$) in quadrature for determining the total uncertainty
\begin{equation}
    \sigma_\text{tot,i}^2 = \sigma_{p,i}^2 + \sigma_\text{sys}^2
\end{equation}
of each integration. This approach serves to fit for underestimated uncertainties in each light curve. For both sampling approaches, we adopted wide uniform priors for all model parameters (see Supplementary Table 1). In the MCMC, we use 500 walkers with 50,000 steps each of which we discard the first 20,000 steps as burn-in and thin the final samples by a factor of 100. The results of all fit parameters' posteriors for both observations' reductions are listed in Supplementary Tables 2, 3, 4 and 5. For the nested samplings, we used 500 live points.

For model comparisons, we calculate all model fits' Bayesian Information Criteria (BIC) from the MCMC samplings and their Bayesian evidences ($Z$) from the nested samplings. To calculate the statistical significances of the rotation-induced radius change delivered by the observations, we then calculate the differences in BIC ($\Delta$BIC) and the normal logarithm of the Bayes factors ($\ln(B)$) of $\mathcal{M}_{\rm{rot}}$ relative to $\mathcal{M}_{\rm{tra}}$ which suggest conclusive detections in all observations and data reductions (see Extended Data Tables \ref{tab:evidence_niriss} and \ref{tab:evidence_nirspec}). We also compare $\mathcal{M}_{\rm{rot}}$ with free $R_2/R_*$ with setting $R_2/R_*=0$, thus comparing two rotational light curve models with either a quadratic or a linear polynomial for the planet-to-star radius ratio as a function of time. In the NIRISS SOSS data, the light curve modeling approach with free $R_2/R_*$ is slightly preferred over the model with $R_2/R_*=0$ with $\ln(B)=1.6$ in the Firefly and $\ln(B)=1.5$ in the Fu reduction. In the NIRSpec G395H data, setting $R_2/R_*=0$ is slightly preferred over fitting for a free $R_2/R_*$ with $\ln(B)=1.9$ in Firefly and $\ln(B)=1.2$ in Eureka!. Thus, there is no indication to prefer either approach over the other \cite{thorngren26} in all observations and data reductions.

\subsubsection*{Spectroscopic light curves}
To investigate the change of WASP-121\,b's transmission spectrum induced by its rotation, we integrated the NIRSpec light curves over bins of 244 pixel columns of the NRS1 and 237 pixel columns of the NRS2 detector to generate 14 spectroscopic light curves. As in the white light curve fit, we fit these light curves simultaneously, adopting the same eccentricity, argument of periapsis and period as before and fitting for $b$ and $a/R_*$ which are shared between all light curves. All other parameters are different for all light curves. Again, we first ran a least-squares fit and sampled the parameters' posterior distributions using an MCMC. In the MCMC, we initiated 500 walkers in a tight Gaussian ball around the least-squares solution and ran 100,000 steps for each walker, discarding the first 50,000 steps as burn-in. The final samples were thinned by a factor of 200.

We applied three modeling approaches to fit the spectroscopic light curves in order to estimate the model assumptions' impacts on the inferred planet radius coefficients $R_i/R_*$. These three approaches were
\begin{enumerate}
    \item $R_2/R_*=0$ and wide uniform priors on the limb darkening coefficients,
    \item $R_2/R_*$ as a free parameter and wide uniform priors on the limb darkening coefficients, and
    \item $R_2/R_*$ as a free parameter and Gaussian priors on the limb darkening coefficients which we adopted from the exotic-LD package \cite{exoticld} using the `stagger' grid of stellar models \cite{staggergrid}.
\end{enumerate}
The MCMCs' posteriors of the planet radius polynomial coefficients from these three modeling approaches analyzing the data from the Firefly reduction and modeling approach number one applied to the Eureka! reduction are plotted in Extended Data Fig. \ref{fig:spectrum}. They reveal that the $R_1/R_*$ coefficient, which gives the linear rate of the planet's radius change relative to the star's radius during transit, is indifferent toward the choice of priors for the limb darkening. This is because $R_1/R_*$ is the coefficient for a polynomial term that is odd about the point of conjunction. As such, a nonzero $R_1/R_*$ will induce a transit shape that is asymmetric about the point of conjunction which cannot be replicated by changing the limb darkening or the orbital parameters. When we adopt $R_2/R_*=0$ in $\mathcal{M}_{\rm{rot}}$, the measurements of $R_1/R_*$ again do not change (see Extended Data Fig. \ref{fig:spectrum}) and thus, our measurements of $R_1/R_*$ are robust against choices of limb darkening priors and the value of $R_2/R_*$.

In contrast to $R_1/R_*$, the measurements reveal that $R_0/R_*$, giving the planet-to-star radius ratio at the point of conjunction, and $R_2/R_*$, describing the curvature of the radius function in time, are both sensitive to modeling approaches of the stellar limb darkening, as their uncertainties greatly increase when moving from tight Gaussian priors to uninformative uniform priors on $u_1$ and $u_2$. The reason for this degeneracy is that $R_2/R_*$ describes an even term about the point of conjunction and is correlated with $R_0/R_*$ (see Extended Data Fig. \ref{fig:corner}). Any even changes about the point of conjunction (such as a decrease of the radius in the first and a converse increase in the second half of the transit) has a similar effect on the light curve as the limb darkening that is symmetric about the center of the stellar disk. Therefore, $R_2/R_*$ and the correlated $R_0/R_*$ cannot be reliably disentangled from the limb darkening using the JWST data.

\subsection*{SPARC/MITgcm predictions}
Synthetic light curves have been generated from the output of a SPARC/MITgcm simulation of WASP-121b \cite{showman09,parmentier18}, post-processed using Pytmosph3R \cite{caldas19,falco22,falco24}. The SPARC/MITgcm solves the primitive equations on a cubic-sphere grid. 
% Parmentier et al. 2013, 2016; Kataria et al. 2015, 2016; Lewis et al. 2017)
In the simulation, the pressure ranges from 200~bar to $2~\mu$bar over 53 levels, with a horizontal resolution equivalent to 128 longitudes and 64 latitudes.
The model includes optical absorbers such as TiO and VO, assumes a solar metallicity and C/O ratio and does not include clouds. As an indication, the temperature on the dayside is $\sim 2800$\,K. 

Pytmosph3R is an open-source Python package that provides tools to compute synthetic observations, transmission and emission spectroscopy, transit light curves and phase curves, from three-dimension atmospheric models such as GCMs.
Its inputs are the planetary, stellar and orbital characteristics, along with the temperature map provided by the GCM, and the abundances of species present in the atmosphere. It computes absorption, Rayleigh and Mie scattering along with continuum absorption (Collision Induced Absorption, CIA) along a set of light rays passing through the atmosphere.
The species included are He, H$_2$, H, H$_2$O, CO, TiO, VO, Na, K, SiO, for which we account for the absorption opacity (except for He, H$_2$ and H) and the continuum includes H$_2$-H$_2$ and H$_2$-He.
Opacities have been downloaded from Exomol \cite{exomol}.
The orbital period has been set to 1.274925~days, with an orbital radius of 0.025~au, a planet radius of 1.61 Jupiter radii, with a surface gravity of 8.436\,m\,s$^{-2}$.

To compare the simulation results with the observations, we generated light curves for WASP-121\,b in NIRSpec G395H's wavelength range with the star's limb darkening set to zero. That way, the depth of the model light curve during the full transit (between contact points 2 and 3) can be converted into the planetary-to-stellar radius ratio using
\begin{equation}
    R_p/R_*(t) = \sqrt{1-F(t)},
\end{equation}
where $F(t)$ is the model light curve's flux. To facilitate the comparison between model light curves and the observations of $R_p/R_*(t)$ approximated as a polynomial (see Eq. \ref{eq:Rp}), we fit quadratic polynomials to the model $R_p/R_*(t)$ of each spectral channel to infer the model predictions for $R_0/R_*$, $R_1/R_*$ and $R_2/R_*$ as functions of wavelength (examples for this approximation are shown in Extended Data Fig. \ref{fig:radius-functions}). All relative deviations of the fit polynomials to the GCM's $R_p/R_*(t)$ are smaller than $0.1\,\%$ in all simulated time steps and thus, the approximation as quadratic functions for the comparison to the observations is justified.

Like the observations, the SPARC/MITgcm shows positive $R_1/R_*$ for most and negative $R_1/R_*$ with much smaller absolute values for some spectroscopic channels (see Extended Data Fig. \ref{fig:spectrum}); however, it systematically underestimates the values for $R_1/R_*$ over the entire wavelength range. One limitation that could contribute to the model's underestimation of $R_1/R_*$ is the lack of clouds in the atmosphere that might be present around the morning terminator. Additionally, a previous phase curve analysis of WASP-121\,b showed that the SPARC/MITgcm delivers too high nightside temperatures compared to the measurements \cite{MikalEvans23}. With an overestimated nightside temperature, the model underestimates the day- to nightside temperature contrast, leading to underestimated longitudinal temperature gradients across the terminators in line with the observed discrepancy to the observations. Another reason for the mismatch between the observations and the model could be an inaccurate atmospheric chemical composition in the model due to the assumed Solar metallicity and $\text{C}/\text{O}=0.55$, since recent observations have consistently shown that WASP-121\,b's C/O is much higher than 0.55 \cite{smith24,pelletier25,EvansSoma25}. These differences in chemistry might also be responsible for the deviations of the model's transmission spectrum at the point of conjunction from the observations (Fig. \ref{fig:spectrum}, blue diamonds in the upper panel), namely, the too steep H$_2$O feature and the underestimation of the CO feature between $4.3\,\mu$m and $5.2\,\mu$m.

To simulate the effect of a colder morning terminator, we modified the `nominal' SPARC/MITgcm by setting the vertical temperature structure from $240^\circ$ to $302^\circ$ in longitude to its structure at $240^\circ$ (see Extended Data Fig. \ref{fig:gcm-maps}). With this modified temperature field (labeled the `muted leading limb' model in Extended Data Fig. \ref{fig:spectrum}), the model predictions for $R_1/R_*$ increase in all wavelengths, because the planet's rotation does not lead to a shrinking of the leading limb anymore due to a vertical temperature structure that is constant in longitude. With this modification, the model predictions for $R_1/R_*$ are larger than the observations between $3.05\,\mu$m and $4.0\,\mu$m where H$_2$O is the dominant molecular absorber \cite{Gapp25}. At wavelengths larger than $4.3\,\mu$m where CO is the strongest absorbing molecule \cite{Gapp25}, the model values for $R_1/R_*$ are smaller than the data (see Extended Data Fig. \ref{fig:spectrum}), though the lower limits of four of the five data points' $1\sigma$ intervals overlap with the model predictions. Thus, the order of magnitude of the transit asymmetry observed with NIRSpec G395H NRS2 can be reproduced by the longitudinal temperature gradient around the evening terminator in the SPARC/MITgcm, when the longitudinal temperature gradient around the morning terminator is muted.

The SPARC/MITgcm predicts negative $R_2/R_*$ through the whole spectrum, suggesting a local maximum of the planet's cross-sectional area during transit. When adopting Gaussian priors on the limb darkening in the light curve analysis, we find consistently positive $R_2/R_*$ (see Extended Data Fig. \ref{fig:spectrum}), implying a local radius minimum during transit. The SPARC/MITgcm does not include any tidal deformation of the planet and thus the local maximum during transit in the model is solely caused by the atmospheric structure. The tidal deformation of WASP-121\,b would lead to an elongation of the planet along the planet-star axis \cite{leconte11} which, together with the planet's rotation, would cause a local minimum of its cross-sectional area during transit \cite{falco24}. Thus, the observed tendency to positive $R_2/R_*$ might be a hint at the planet's tidal deformation. Measuring the planet's tidal deformation from this observation would be dependent on the assumption that the adopted limb darkening priors are accurate. However, their degree of precision is unknown and thus, we refrain from attempting to constrain WASP-121\,b's tidal Love number that quantifies its tidal deformation.

\subsection*{Other sources of asymmetric transits}
A phase-dependent transmission spectrum of WASP-121\,b is not the only possible cause of asymmetric transit light curves. Thus, we examine the plausibility of the observed transit asymmetries originating in the observing instruments or the planet's host star WASP-121\,A.

\subsubsection*{Instrumental systematics}
Instrumental systematics causing the observed asymmetric light curve shapes would require an instrumental effect that impacts the NIRISS SOSS and NIRSpec G395H's NRS2 data that show a transit asymmetry, but not NIRSpec G395H's NRS1 observations that deliver a symmetric transit light curve (see Fig. \ref{fig:white-lightcurve} and Extended Data Fig. \ref{fig:white-lightcurve_eureka}). Considering that past observations suggested the NRS2 detector to be less affected by instrumental systematics than NRS1 (see, e.g., ref. \cite{MikalEvans23,wallack24,alderson25}) and the coincidence of systematics required during the precise time span of the transit to create the observed signal in both the NIRISS SOSS and the NIRSpec G395H observations, such a scenario appears unlikely. Therefore, the transit asymmetry is likely a signal from either WASP-121\,b or its host star.

\subsubsection*{Stellar activity and gravity darkening}
In NIRSpec's wavelength range, the observed transit asymmetry is more pronounced in the longer than in the shorter observed wavelengths. This is the opposite of how transit asymmetries caused by temperature inhomogeneities on the star's surface driven by, e.g., star spots or faculae would change with wavelength \cite{tls1,kostogryz25}. Additionally, as an F6-type star \cite{delrez16}, WASP-121\,A is expected to display less activity-driven flux variations throughout its surface that could induce asymmetric transit light curves than later stellar types \cite{tls2}. Indeed, previous transit observations in the optical have not revealed any hints for photometric contamination caused by stellar activity \cite{Evans18}, making stellar activity as the cause of the observed transit asymmetries yet more unlikely. As an early spectral type, however, WASP-121\,A might be a fast rotator, leading to gravity darkening of its equator relative to the poles due to the reduction of the equator's surface gravity caused by centrifugal forces and an oblate shape of the star \cite{vonzeipel24}. Together with the misalignment of WASP-121\,b's orbital axis with the star's rotation axis \cite{delrez16,bourrier2020a}, this system configuration can induce a transit asymmetry \cite{barnes09}.

From the nearly pole-on inclination of WASP-121\,A \cite{bourrier2020a}, we estimate the rotation rate of the star to be between $0.15-0.28$ as a fraction of the critical rotational velocity. As such, gravity darkening of the host star should be a significant effect, causing a difference in emission intensity between the equator and pole of $6.4-20.8\%$ in the optical and $2-6\%$ in the JWST/NIRSpec passband. To quantify the possible impact of gravity darkening on our observations, we analyzed all available observations of WASP-121\,b observed with TESS due to the proportionally higher contribution of gravity darkening to the transit asymmetry in the optical. We query 5 sectors of 20\,s cadence light curves and 1 sector of 120\,s cadence, including only data $\sim 0.25$\,days around the center of each transit. To model the transit of a rapidly-rotating, gravity-darkened star, we use the method described in ref. \cite{dholakia22}, but instead rewrite the spherical-harmonic decomposition of the gravity-darkened surface using Planck's blackbody law into JAX. This allows us to use \texttt{eclipsoid} \cite{dholakia25} as the model within the JAX ecosystem. We use NumPyro \cite{phan2019} to construct a probabilistic model for the gravity-darkened transit, using priors from ref. \cite{MikalEvans23} for the stellar and planet parameters and ref. \cite{bourrier2020a} for the projected orbital obliquity. We also sample in the sine of stellar inclination as an isotropic prior on the spin axis, restricted to the range 0-0.25 to ensure the star is relatively pole-on, as suggested by ref. \cite{bourrier2020a}, and place a uniform prior on the dimensionless stellar rotation rate $\omega$ between 0-0.3. We fix the gravity-darkening exponent $\beta$ that governs what temperature difference a difference in surface gravity between the star's equator and pole results in to 0.23 as varying it can cause it to run off towards physically unrealistic values \cite{ahlers20b}. Finally, we use Hamiltonian Monte Carlo with the NUTS sampler \cite{hoffman2011} in order to obtain posteriors on the parameters of interest. We also performed inference using the same priors but without the gravity-darkening map on the star as a comparison to a standard transit fit to examine the amplitude of a transit asymmetry the star's gravity darkening induces.

TESS's light curves confirm that WASP-121\,A is measurably gravity-darkened, creating a transit asymmetry with an amplitude on the order of 100\,ppm in the optical light curves between contact points 2 and 3 as well as upward residuals reaching up to $\sim 300$\,ppm during egress (see Extended Data Fig. \ref{fig:gravitydarkening}). The posteriors of the planet's and star's inclination as well as the projected and true spin-orbit angles (see Supplementary Table 6) are consistent with previous radial-velocity observations of the system \cite{bourrier2020a}. For the stellar rotation rate, we find $(12.6\pm 0.19)\,\%$ times the critical rotational velocity which is slightly lower than the $(15-28)\,\%$ estimated before \cite{bourrier2020a} and thus suggests that the star's rotation rate lies on the lower end of previous constraints. One possible reason for the small discrepancy between our result and the radial-velocity observation's lower limit might be the star's gravity-darkening exponent we fixed at $\beta=0.23$. $\beta$ can be lower than the value of $\beta=0.25$ predicted theoretically \cite{monnier07}, reducing the temperature contrast between stellar equator and pole for a given rotation rate. Thus, if WASP-121\,A's $\beta$ were lesser than $0.23$, our measured $\omega$ would be biased to smaller values which would be in line with our measurement being slightly lower than the Doppler tomography's lower limit.

Since the wavelength bands of TESS and JWST/NIRISS SOSS Order 1 overlap, the NIRISS SOSS light curve will also be affected by gravity darkening. Indeed, both the asymmetry between contact points 2 and 3 and the yet stronger upward residuals during egress caused by gravity darkening in TESS (see Extended Data Fig. \ref{fig:gravitydarkening}) are clearly visible in the white light curve of NIRISS SOSS Order 1 (see middle rows of Fig. \ref{fig:white-lightcurve} and Extended Data Fig. \ref{fig:white-lightcurve_eureka}).

To examine, if the transit asymmetry observed with JWST/NIRSpec G395H's NRS2 is consistent with WASP-121\,A's gravity darkening observed with TESS, we fit the JWST/NIRSpec G395H NRS2 data in the same manner as before, but with a maximum a-posteriors method. The stellar rotation rate required to fit the NRS2 light curve is $\omega=19.8\,\%$ (see Supplementary Table 6) and thus, deviates from the TESS result by $>30\sigma$. This demonstrates that gravity darkening of WASP-121\,A is not sufficient for explaining the transit asymmetry observed with JWST/NIRSpec. Additionally, the gravity darkening model does not offer an explanation for the lack of a transit asymmetry in NRS1, if NRS2 were affected by gravity darkening as the transit asymmetry driven by gravity darkening would monotonically decreases as longer wavelengths are observed \cite{barnes09}. Therefore, in NRS2, gravity darkening must be a minor effect compared to WASP-121\,b's phase-dependent transmission spectrum as the symmetric NRS1 light curve demonstrates that gravity darkening has already dropped below a measurable level at that wavelength range (see Fig. \ref{fig:white-lightcurve} and Extended Data Fig. \ref{fig:white-lightcurve_eureka}).

\subsection*{Phase-dependent CO absorption}
JWST/NIRISS’s and JWST/NIRSpec G395H’s bandpasses include one CO band each. To examine, if WASP-121\,b’s radius change is consistent between both observed CO bands, we integrated both detectors’ light curves over the heads of the two CO bands and fit both $\mathcal{M}_{\rm{tra}}$ and $\mathcal{M}_{\rm{rot}}$ with $R_2/R_*=0$ using least-squares approaches and MCMCs as before to the light curves. In these fits, the orbital parameters were fixed to the two observations’ white light curve fit results. The fits to the observations (see Extended Data Fig. \ref{fig:co-lightcurve}) reveal that WASP-121\,b's rotation-induced radius change results in residuals on the order of 250 ppm between contact points 2 and 3 when $\mathcal{M}_{\rm{tra}}$ is used in both the NIRISS and NIRSpec light curve, though the NIRISS data suffer from a lower signal-to-noise ratio due to the probed CO band being narrower than the one in NIRSpec. The MCMC posteriors of $R_1/R_*$ are $650^{+188}_{-189}$\,ppm\,hr$^{-1}$ in NIRISS and $580^{+90}_{-91}$\,ppm\,hr$^{-1}$ in NIRSpec and thus consistent between both observations.

The slightly stronger increase of WASP-121\,b's apparent radius as a function of time in NIRISS compared to NIRSpec might be the result of contamination with gravity darkening. We note, however, that the linear rate of radius change in the white light curve of NIRISS ($409^{+53}_{-53}$\,ppm\,hr$^{-1}$, see Supplementary Table 4) is smaller than the rate measured from the light curve integrated over the CO band ($650^{+188}_{-189}$\,ppm\,hr$^{-1}$). Gravity darkening monotonically decreases with wavelength and since the CO band in NIRISS is located on the long-wavelength end of the detector, there has to be an additional source of an asymmetric light curve that elevates the $R_1/R_*$ measurement over that from the white light curve. Thus, both the NIRISS and the NIRSpec data point toward increasing CO absorption in the planet as a function of orbital phase with quantitatively consistent amplitudes. This demonstrates JWST's remarkable capability to measure phase-dependent absorption in transiting exoplanets consistently between different observations and instruments.

\section*{Data availability}
The JWST and TESS data used in this study are publicly available on the Barbara A. Mikulski Archive for Space Telescopes at \url{https://mast.stsci.edu}. The JWST/NIRSpec observation analyzed here is part of JWST observing program G.O. 1729 (P.I.: T.M.E.-S.; co-P.I.: Tiffany Kataria) and can be retrieved from ref. \cite{nirspec_doi}. The JWST/NIRISS observation is associated with JWST observing program G.T.O. 1201 (P.I.: David Lafreni\`ere) and can be retrieved from ref. \cite{niriss_doi}.

\section*{Code availability}
The following software packages were used to perform the data analysis and model interpretation: astropy \cite{astropy13,astropy18,astropy22}, batman \cite{batman}, corner \cite{corner}, dynesty \cite{dynesty}, eclipsoid \cite{dholakia25}, emcee \cite{emcee}, Eureka! \cite{eureka}, ExoTIC-LD \cite{exoticld}, FIREFly \cite{rustamkulov22,rustamkulov23,Sing2024Natur.630..831S}, lmfit \cite{lmfit}, matplotlib \cite{matplotlib}, numpy \cite{numpy}, NumPyro \cite{phan2019}, pandas \cite{pandas}, scipy \cite{scipy}

\section*{Acknowledgements}
This work is based on observations made with the NASA/ESA/CSA James Webb Space Telescope. The data were obtained from the Mikulski Archive for Space Telescopes at the Space Telescope Science Institute, which is operated by the Association of Universities for Research in Astronomy, Inc., under NASA contract NAS 5-03127 for JWST.

This research has made use of the NASA Exoplanet Archive, which is operated by the California Institute of Technology, under contract with the National Aeronautics and Space Administration under the Exoplanet Exploration Program.

\section*{Author contributions}
C.G. analyzed the JWST data and wrote the manuscript with assistance from A.F, T.M.E.-S., D.K.S, S.D. and E.-M.A. A.F. provided model light curves derived from the SPARC/MITgcm using Pytmosph3R. T.M.E.-S., V.P. and J.L. supported the interpretation of the results. D.K.S and E.-M.A. reduced the JWST/NIRSpec observations. D.K.S and G.F. reduced the JWST/NIRISS observations. S.D. analyzed the TESS data and the JWST/NIRSpec NRS2 data with a gravity darkening model.

\section*{Competing interests}
The authors declare no competing interests.

\section*{Materials and correspondence}
Correspondence and requests for materials should be addressed to Cyril Gapp.

\bibliography{references}

\begin{appendices}
\section*{Extended Data}
\renewcommand\figurename{Extended Data Fig.}
\renewcommand\tablename{Extended Data Table}
\begin{figure}[h]
    \centering
    \includegraphics[width=1.0\textwidth]{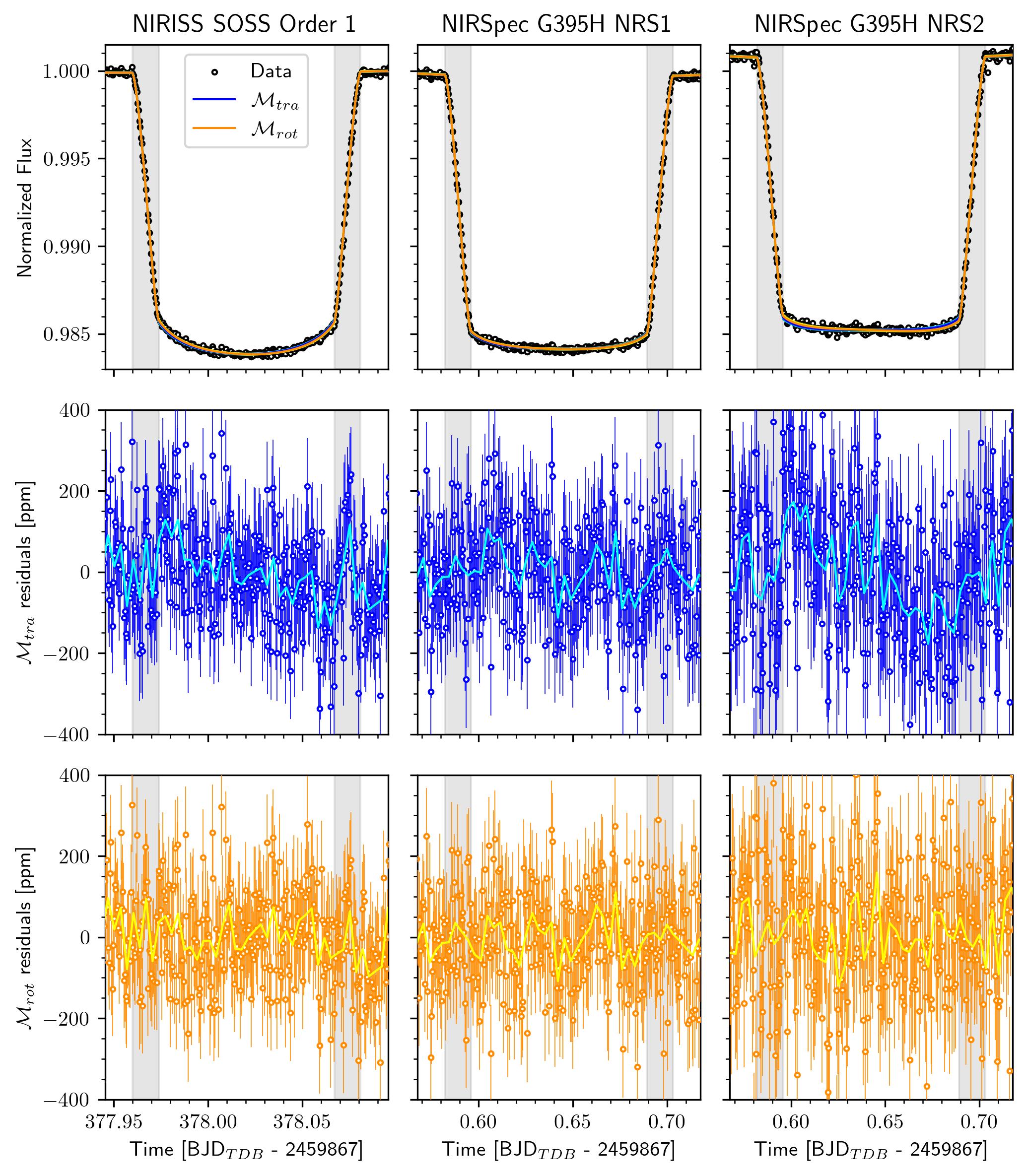}
    \caption{WASP-121\,b’s white transit light curves observed with JWST/NIRISS SOSS and JWST/NIRSpec G395H. The same as Fig. \ref{fig:white-lightcurve}, but using the Fu reduction of the JWST/NIRISS observations and the Eureka! reduction of the JWST/NIRSpec observations.}
    \label{fig:white-lightcurve_eureka}
\end{figure}
\begin{figure}[h]
    \centering
    \includegraphics[width=1.0\textwidth]{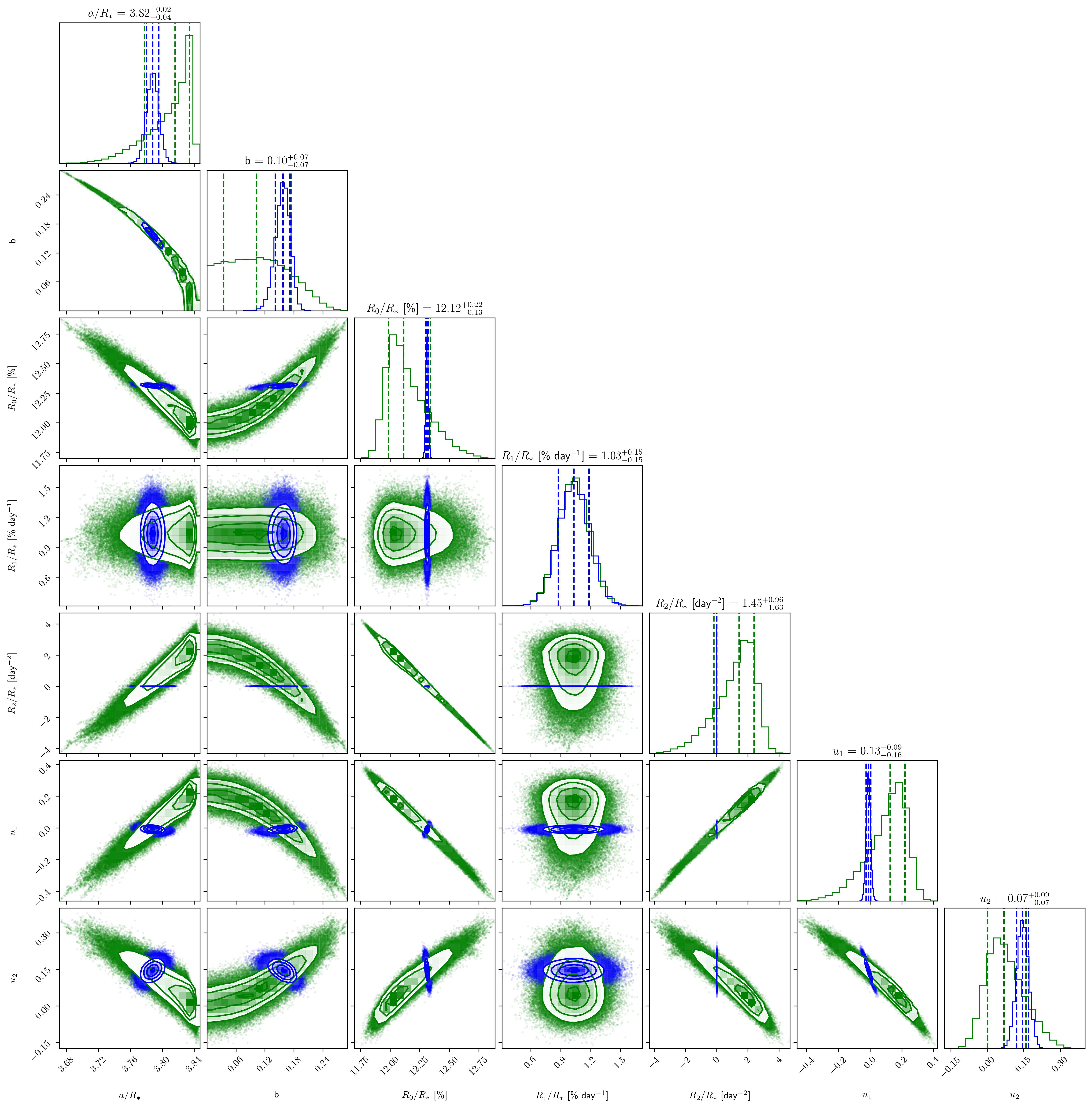}
   \caption{Posteriors of selected fit parameters from the MCMC and using $\mathcal{M}_{\rm{rot}}$ to NIRSpec's white light curves reduced with Firefly. Blue contours show the posteriors from the light curve fit with $R_2/R_*=0$ and green contours show the posteriors with free $R_2/R_*$. The orbital parameters ($a/R_*$ and $b$) were shared between both detectors' light curves and all other plotted parameters correspond to those that were fitted to the NRS2 light curve only. For the latter, equivalent parameters were fitted to the NRS1 light curve but are not shown for visual clarity. The values in the panel titles show the median and $1\sigma$ interval of the fit parameters from the light curve fit with free $R_2/R_*$.}
  \label{fig:corner}
\end{figure}
\begin{figure}[h]
    \centering
    \includegraphics[width=1.0\textwidth]{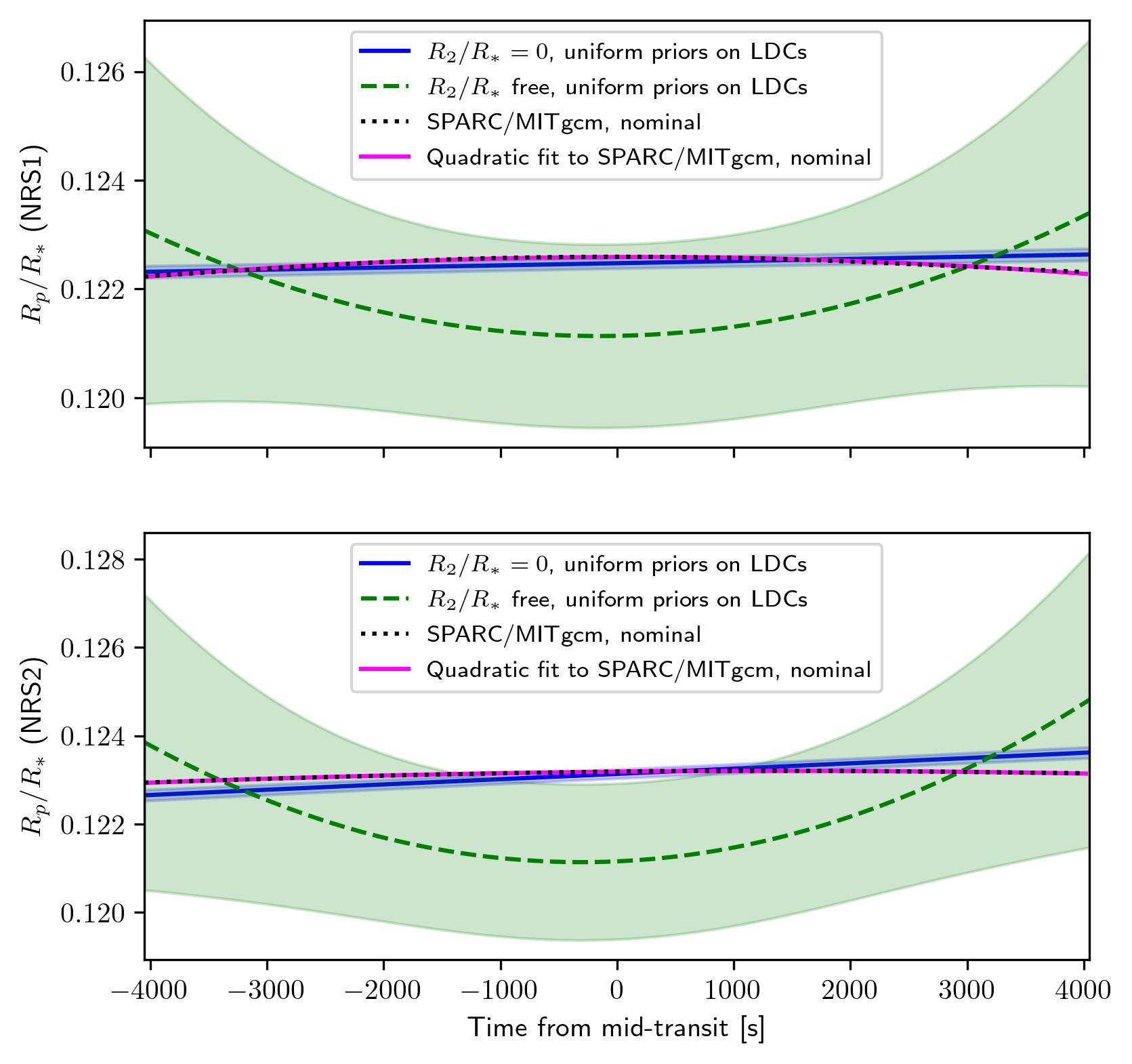}
   \caption{WASP-121\,b’s transit radius as a function of time observed with JWST/NIRSpec and simulated using SPARC/MITgcm. The upper and lower panel show the observations and model for the white light curves of NIRSpec's NRS1 and NRS2 detectors, respectively. Lines show the median measurements with the 1\,$\sigma$ intervals of the measurements indicated using colored shadings. In addition to the model result shown with a dotted line, we also show a quadratic fit to that model using a solid magenta line. Any deviations of the quadratic fit from the model are well below the measurement uncertainties.}
  \label{fig:radius-functions}
\end{figure}
\begin{figure}[h]
    \centering
    \includegraphics[width=1.0\textwidth]{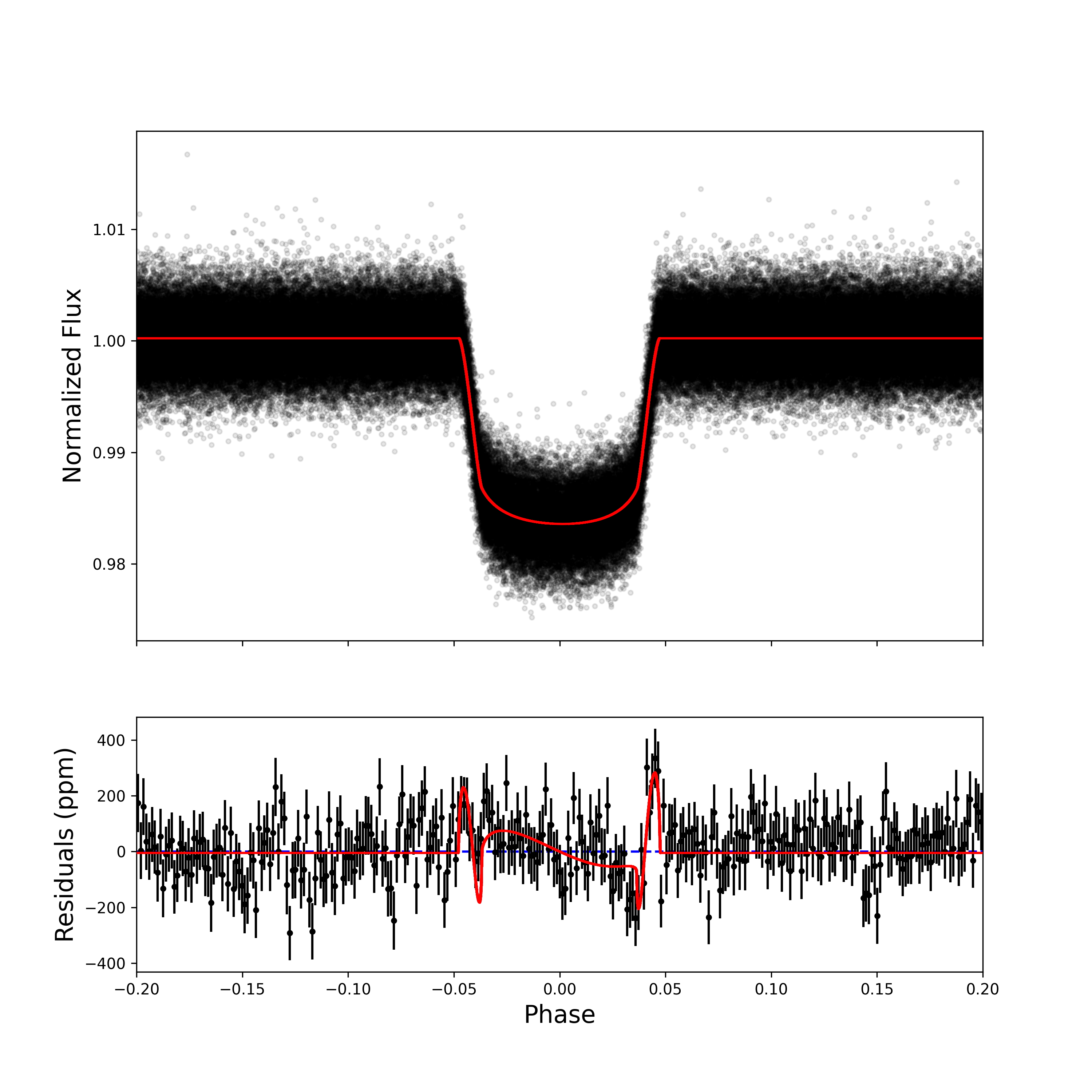}
    \caption{TESS optical light curve of the transit of WASP-121\,b showing asymmetry due to gravity-darkening of the rapidly rotating host star. The top panel shows the unbinned, phased TESS light curve centered around the transit, along with the gravity-darkened model overplotted in red. The bottom panel shows the binned residuals from a maximum a-posteriori (MAP) fit to a simple limb-darkened star without gravity darkening. Overplotted in red is the gravity-darkened fit. The error bars are $1\sigma$ uncertainties.}
    \label{fig:gravitydarkening}
\end{figure}
\begin{figure}[h]
    \centering
    \includegraphics[width=1.0\textwidth]{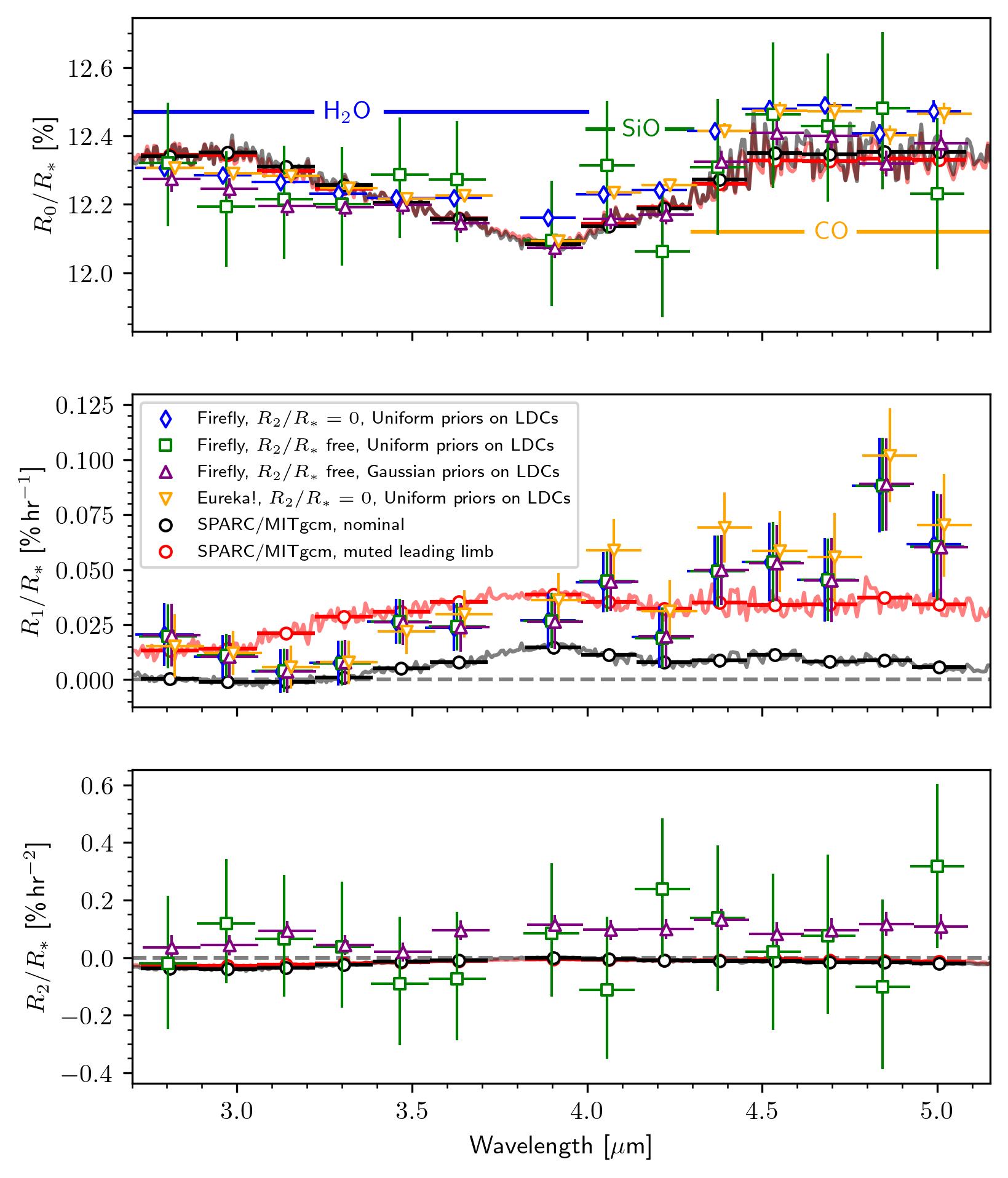}
    \caption{WASP-121\,b's radius polynomial coefficients (see Eq. \ref{eq:Rp}) as functions of wavelength measured using the JWST/NIRSpec G395H observations and modeled using SPARC/MITgcm. Colored diamonds, squares and triangles indicate the medians of 125,000 posterior samples from the spectroscopic light curve fits with different assumptions about $R_p/R_*(t)$ and modeling approaches for the limb darkening coefficients (LDCs) and using one of the two data reductions. All markers' horizontal bars indicate the wavelength ranges of the spectroscopic light curve channels and vertical error bars show the $1\,\sigma$ range of the 150,000 posterior samples. The data were slightly offset to each other horizontally for increased visibility. Solid lines show the model predictions at a spectral resolution of $R\sim 600$ and dots represent the model values binned into the same wavelength bins as the observations. In the upper panel, the model spectra were subtracted by a constant value so that their and Firefly's $R_2/R_*=0$ observations' mean between $2.7$ and $4.0\,\mu$m match. Dashed grey lines at zero were added to guide the eye.}
    \label{fig:spectrum}
\end{figure}
\begin{figure}[h]
    \centering
    \includegraphics[width=1.0\textwidth]{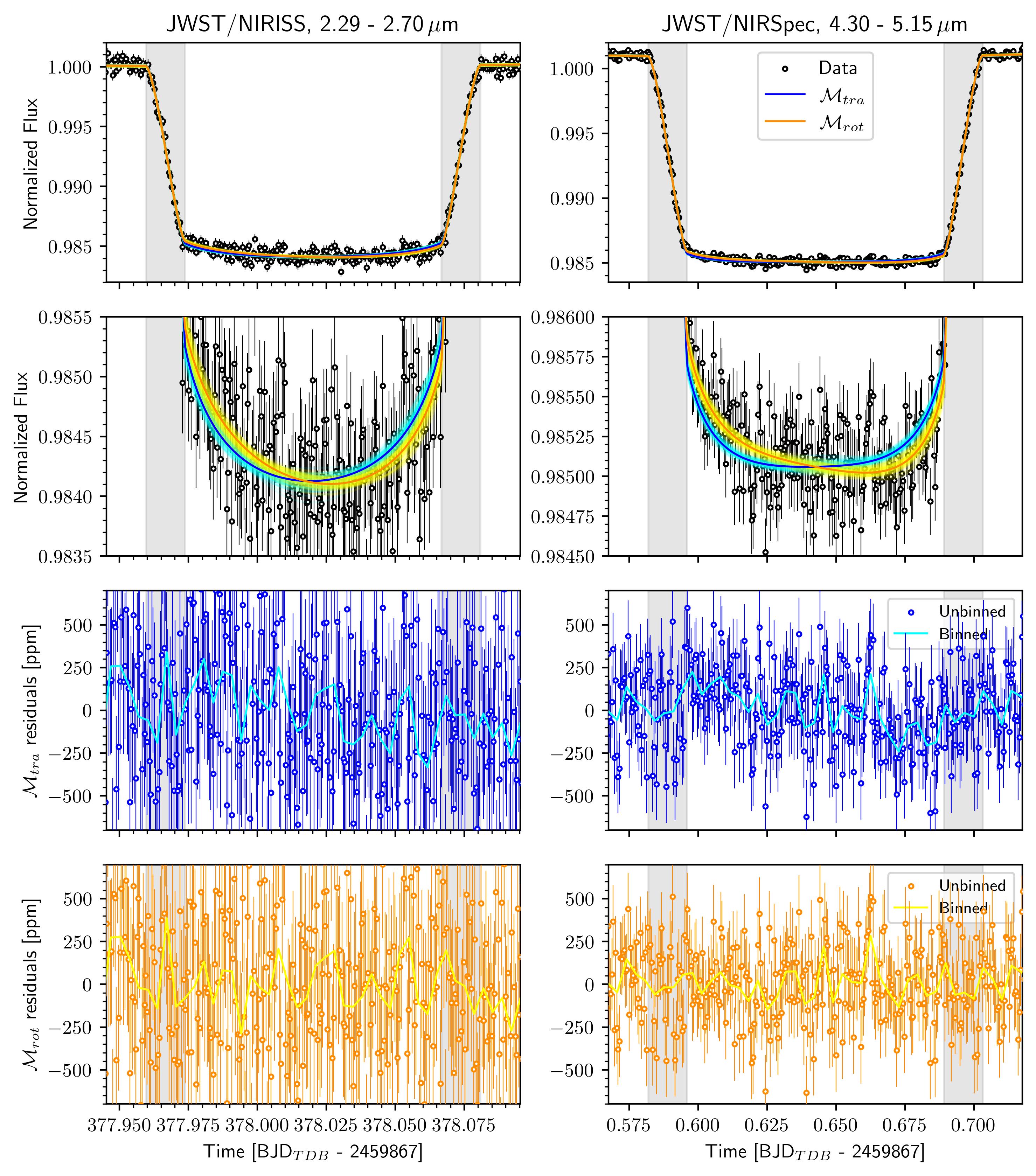}
    \caption{Light curves integrated over the strongest CO bands in NIRISS and NIRSpec. Both observations were reduced using Firefly. The first two rows show the raw light curves with the maximum-likelihood translational ($\mathcal{M}_{\rm{tra}}$) and rotational ($\mathcal{M}_{\rm{rot}}$) models inferred from a Markov-Chain Monte Carlo (MCMC) sampling of the posterior probability. The second row shows a zoom into the bottom of the transit. Circles show the fluxes per integration normalized by the median fluxes measured during the secondary eclipse observed before the transit and error bars depict the $1\sigma$ intervals measured from 150,000 posteriors samples of the light curve. The transparent cyan and yellow lines depict models calculated from 100 samples randomly drawn from the MCMC chains of $\mathcal{M}_{\rm{tra}}$ and $\mathcal{M}_{\rm{rot}}$, respectively. The third and fourth rows show the residuals between the maximum-likelihood models and the data. Vertical gray lines mark contact points 1, 2, 3, and 4 calculated from the planet's orbital parameters and radius \cite{winn10} with gray shaded regions indicating the times of ingress and egress.}
    \label{fig:co-lightcurve}
\end{figure}
\begin{figure}[h]
    \centering
    \includegraphics[width=1.0\textwidth]{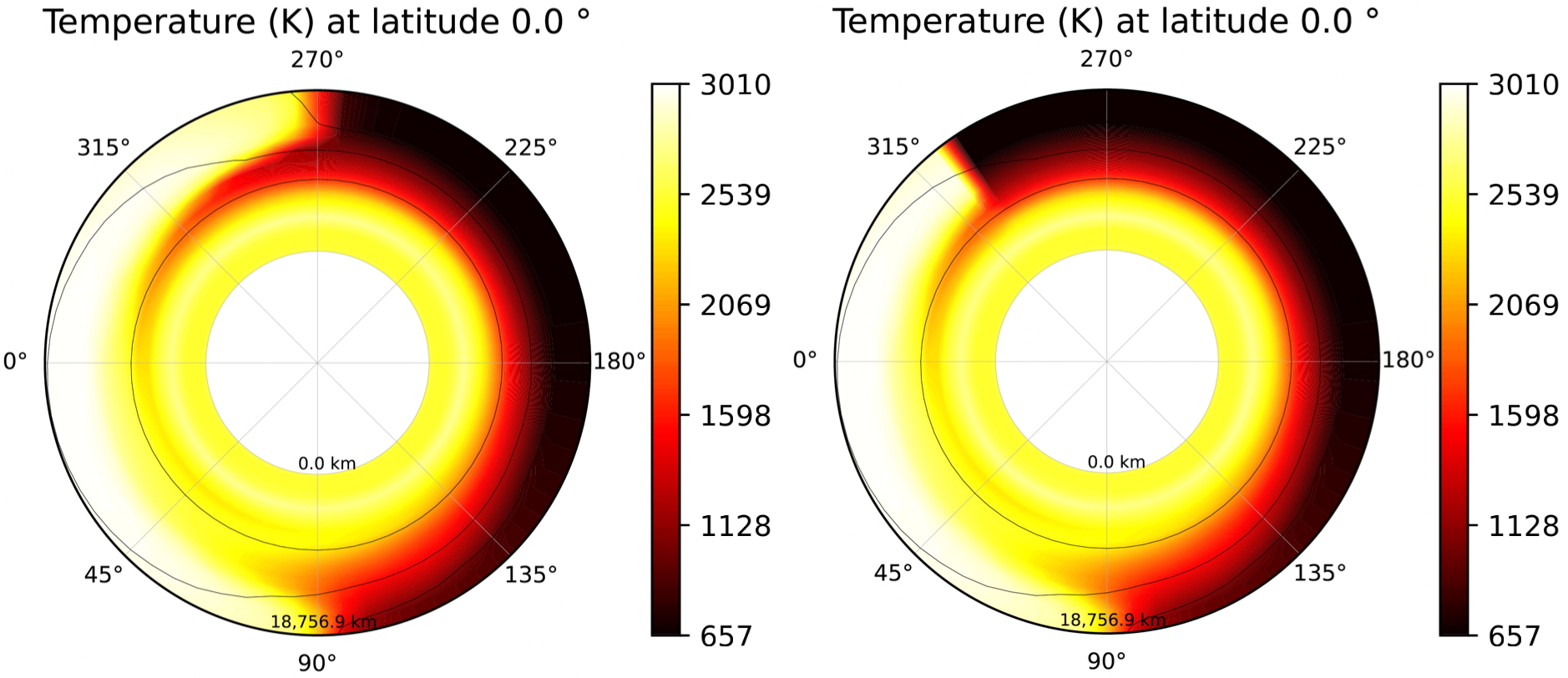}
    \caption{Equatorial cross-sections of WASP-121\,b's temperature field in the SPARC/MITgcm. The left panel shows the nominal temperature field and the right panel shows the temperature field of the `muted leading limb' case. The modified temperature field is identical to the nominal case, but has the vertical temperature profile from $240^\circ$ to $302^\circ$ in longitude set to its structure at $240^\circ$. The atmosphere has been enlarged compared to the planet core for better visibility.}
    \label{fig:gcm-maps}
\end{figure}
\begin{table}[h]
    \caption{Metrics from the MCMC and nested samplings to the NIRISS SOSS first order white light curve.}
    \begin{tabular}{@{}l|ccc|ccc@{}}
    \toprule
    \textbf{Data reduction} & \multicolumn{3}{c}{\textbf{Firefly}} & \multicolumn{3}{c}{\textbf{Fu}} \\
    \textbf{Model} & $\mathcal{M}_{\rm{tra}}$ & $\mathcal{M}_{\rm{rot}}$ & $\mathcal{M}_{\rm{rot}}$ & $\mathcal{M}_{\rm{tra}}$ & $\mathcal{M}_{\rm{rot}}$ & $\mathcal{M}_{\rm{rot}}$ \\
    & & $R_2/R_*$ free & $R_2/R_*=0$ & & $R_2/R_*$ free & $R_2/R_*=0$ \\
    \midrule
	\textit{MCMC sampling} \\
	BIC & -15356.39 & -15419.00 & -15415.50 & -15373.82 & -15422.91 & -15418.24 \\
	$\Delta$BIC\footnotemark[1] & & 62.61 & 59.11 & & 49.09 & 44.42 \\
	\textit{Nested sampling} \\
	$\ln(Z)$ & 7651.7 & 7676.6 & 7674.9 & 7659.4 & 7677.3 & 7675.8 \\
	$\ln(B)$\footnotemark[1] & & 24.8 & 23.2 & & 18.0 & 16.5 \\
    \botrule
    \end{tabular}
    \footnotetext[1]{Differences in Bayesian Information Criterion ($\Delta$BIC) and natural logarithms of Bayes factors ($\ln(B)$) are all relative to $\mathcal{M}_{\rm{tra}}$.}
    \label{tab:evidence_niriss}
\end{table}
\begin{table}[h]
    \caption{Metrics from the MCMC and nested samplings, analyzing the white light curves from NIRSpec G395H NRS1 and NRS2 simultaneously.}
    \begin{tabular}{@{}l|ccc|ccc@{}}
    \toprule
    \textbf{Data reduction} & \multicolumn{3}{c}{\textbf{Firefly}} & \multicolumn{3}{c}{\textbf{Eureka!}} \\
    \textbf{Model} & $\mathcal{M}_{\rm{tra}}$ & $\mathcal{M}_{\rm{rot}}$ & $\mathcal{M}_{\rm{rot}}$ & $\mathcal{M}_{\rm{tra}}$ & $\mathcal{M}_{\rm{rot}}$ & $\mathcal{M}_{\rm{rot}}$ \\
    & & $R_2/R_*$ free & $R_2/R_*=0$ & & $R_2/R_*$ free & $R_2/R_*=0$ \\
    \midrule
	\textit{MCMC sampling} \\
	BIC & -30417.62 & -30440.52 & -30453.45 & -29958.45 & -29999.89 & -30013.13 \\
	$\Delta$BIC\footnotemark[1] & & 22.90 & 35.83 & & 41.44 & 54.68 \\
	\textit{Nested sampling} \\
	$\ln(Z)$ & 15154.8 & 15166.5 & 15168.4 & 14924.4 & 14948.1 & 14949.3 \\
	$\ln(B)$\footnotemark[1] & & 11.7 & 13.6 & & 23.7 & 24.9 \\
    \botrule
    \end{tabular}
    \footnotetext[1]{Differences in Bayesian Information Criterion ($\Delta$BIC) and natural logarithms of Bayes factors ($\ln(B)$) are all relative to $\mathcal{M}_{\rm{tra}}$.}
    \label{tab:evidence_nirspec}
\end{table}

\end{appendices}

\end{document}